\documentclass{rspublic}
\usepackage{epsfig}

\def\ltsima{$\; \buildrel < \over \sim \;$}
\def\gtsima{$\; \buildrel > \over \sim \;$}
\def\simlt{\lower.5ex\hbox{\ltsima}}
\def\simgt{\lower.5ex\hbox{\gtsima}}

\begin{document}

\title[Obscured AGN: the hidden side of the X--ray Universe]{
Obscured AGN: the hidden side of the X--ray Universe}

\author[G. Matt]{Giorgio Matt}

\affiliation{Physics Dept., University ``Roma Tre", Roma, Italy}

\label{firstpage}

\maketitle

\begin{abstract}{X-ray Astronomy; Active Galactic Nuclei}

Most Active Galactic Nuclei (AGN) are `obscured', i.e. the 
nucleus is hiding behind a screen of absorbing material. The 
advantage of having the nucleus obscured is to make easier
the observations of those emission components
which originate in circumnuclear matter outside the absorbing regions,
because in this case they are not outshined by the nuclear emission.
This is particularly important in X--rays, where spatial resolution is
(with the notable exception of $Chandra$) poorer than in the optical,
and the study of circumnuclear regions is often based on spectral
analysis only. 

The properties of circumnuclear matter, in the light of 
recent high spectral and/or angular resolution $Chandra$ and
XMM--Newton observations, will be reviewed in Sec.~2. In Sec.~3 we will
discuss obscured AGN in the framework of the Unification Model.
Recent discoveries of X--ray obscured Seyfert 1, and of X--ray
loud but optically normal galaxies,  are calling for a revision of
the  Unification Model.

Obscured AGN have also a cosmological relevance. Not only are they a 
fundamental ingredient of synthesis models of the Cosmic X--ray
Background (XRB),  but provide a link between the XRB and the Cosmic
Infrared Background, as briefly discussed in Sec.~4.

\end{abstract}

\section{Introduction}

\begin{figure}
\epsfig{file=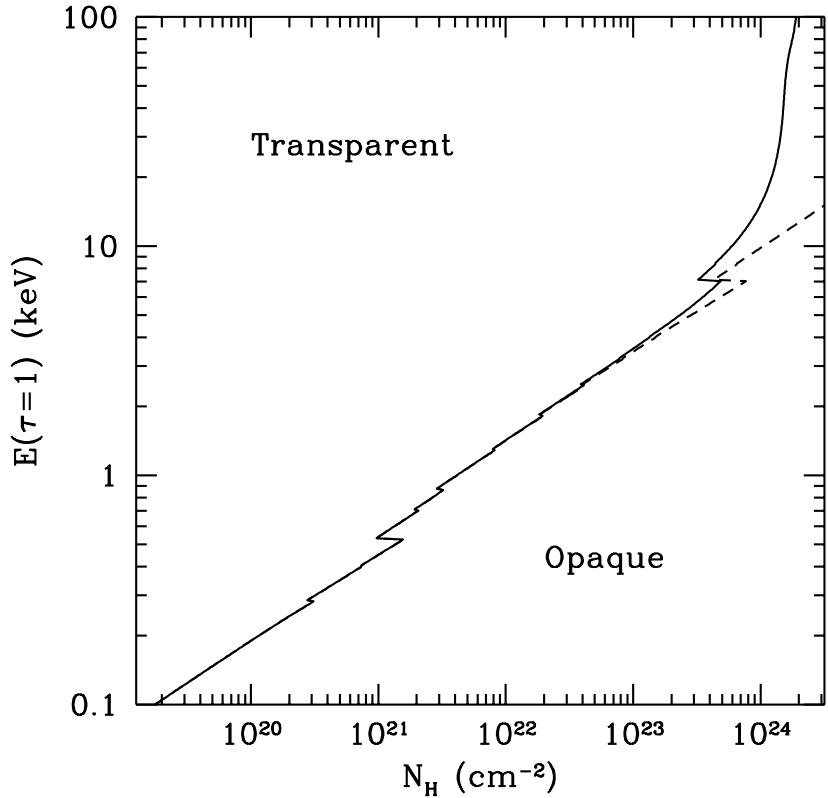, height=12.0cm, width=12.0cm}
\caption{The energy corresponding to $\tau$=1 as a function of the 
column density of the absorbing
matter. The dashed line refers to photoabsorption only, 
the solid line includes Compton scattering.}
\label{nh}
\end{figure}

Both direct (surveys) and indirect (synthesis models of the Cosmic
X--ray Background, CXRB) methods
clearly indicate that most AGN are `obscured' in X--rays, i.e. that
their nucleus is hidden behind an absorbing material, which prevents
the nuclear emission from being directly observed up to the energy (if any,
see below) at which the material becomes transparent
(this energy depending of course
on the column density of the absorber, see Fig.~\ref{nh}). 
Indeed, the three closest 
AGN (Circinus Galaxy, NCG~4945 and Centarus A) are all heavily obscured.
The absorbing matter is often 
very thick, its column density exceeding the value, 
$\sigma_T^{-1}$=1.5$\times$10$^{-24}$
cm$^{-2}$, for which the Compton scattering optical depth equals unity
(in these cases, the sources are called `Compton--thick'. If the column 
density is smaller than $\sigma_T^{-1}$ but still in excess of the 
Galactic one, the source is called `Compton--thin'). Compton--thick
sources provide the most favourable case 
for studying the circumnuclear matter, because 
the emission from this matter (arising from reflection and reprocessing of the
nuclear radiation) is not significantly diluted by the primary emission not only
in soft X--rays but up to 10 keV at least.

\begin{figure}
\begin{minipage}{60mm}
\epsfig{file=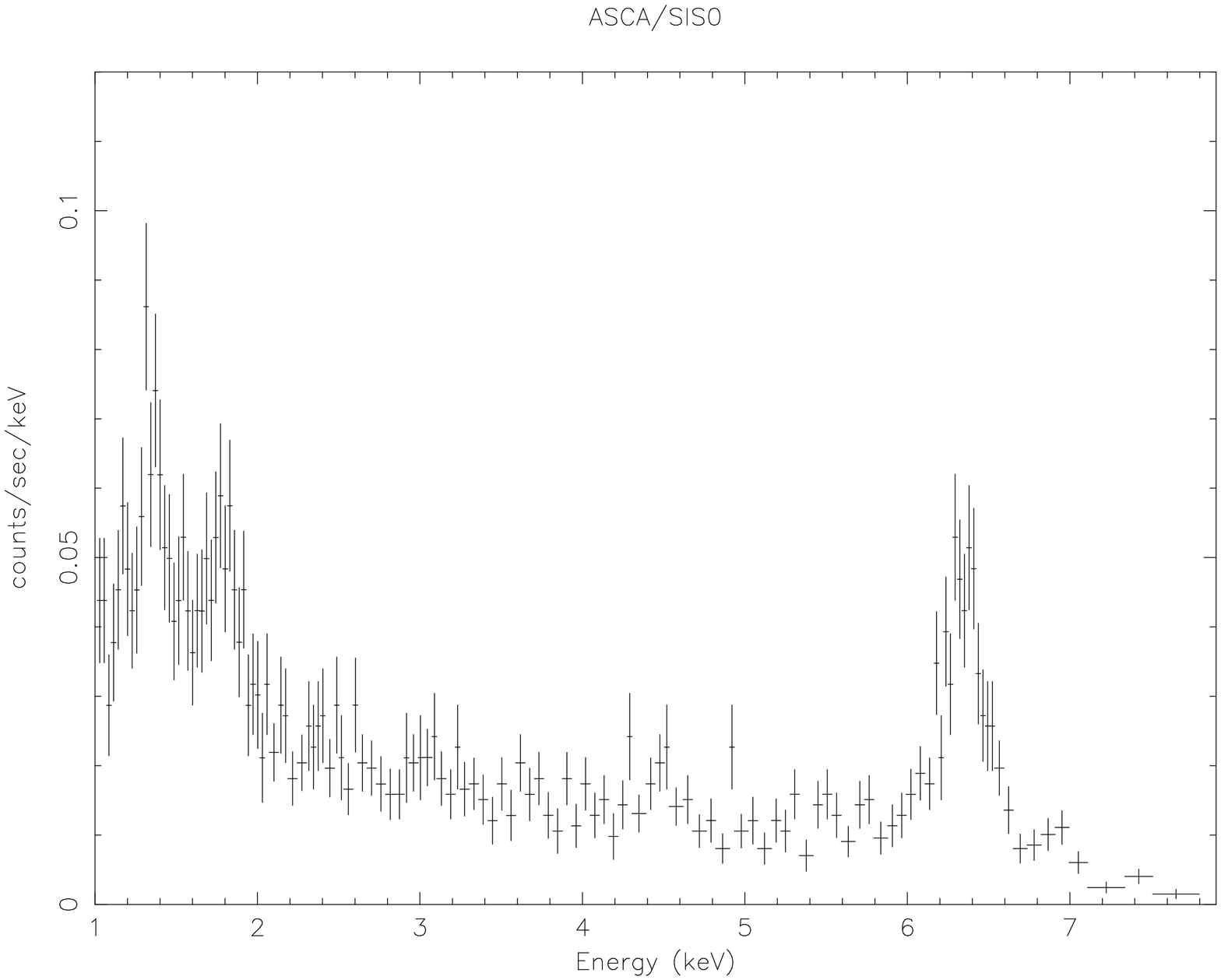,width=65mm, height=65mm, angle=-90}
\end{minipage}
\hspace{\fill}
\hspace{\fill}
\hspace{\fill}
\hspace{\fill}
\begin{minipage}{60mm}
\epsfig{file=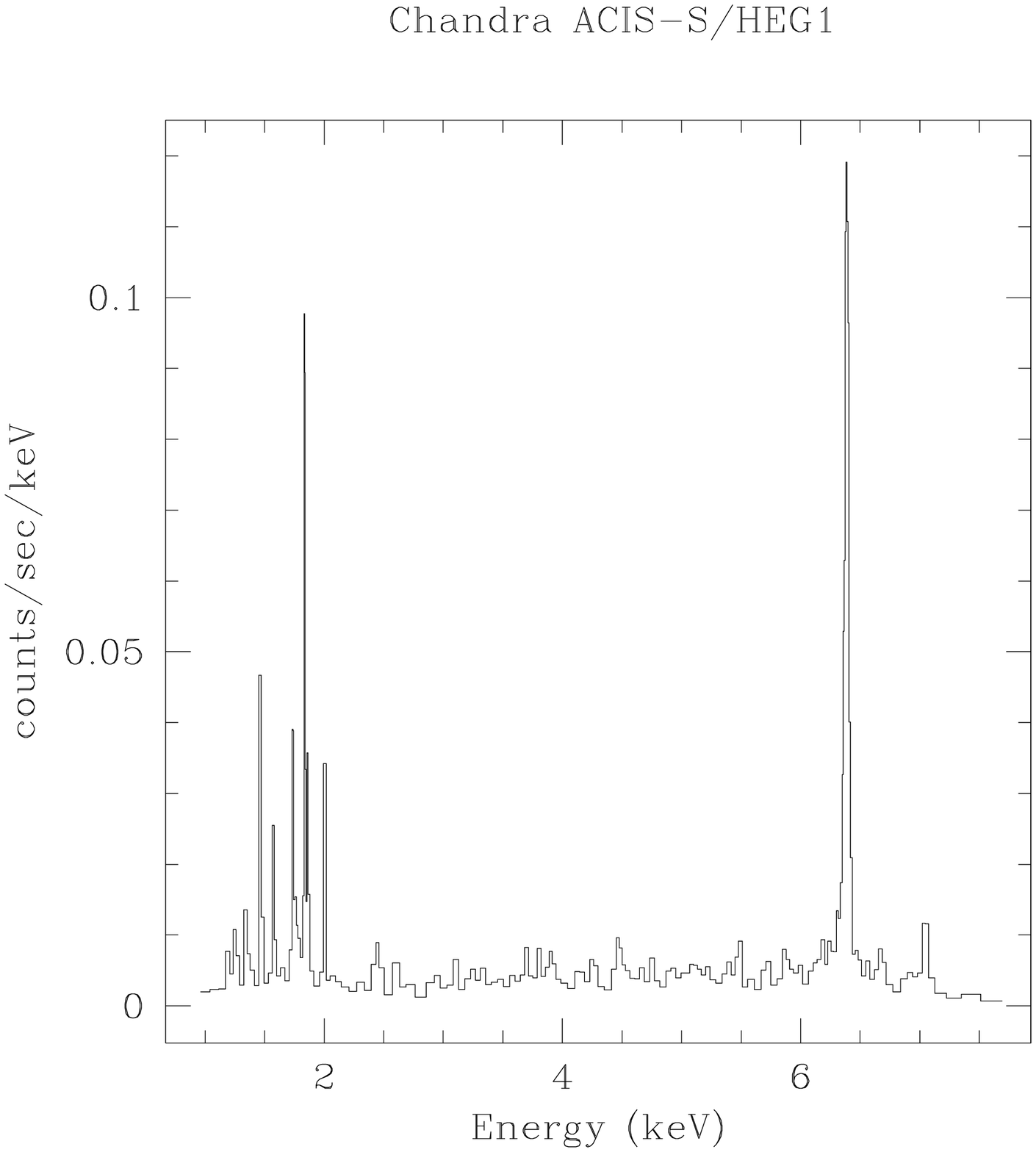,width=65mm, height=65mm}
\end{minipage}
\caption{CCD (ASCA/SIS, left panel) and grating ($Chandra$/HEG, right panel)
spectra of the Circinus Galaxy. More details in Matt et al. (1996a)
and Sambruna et al. (2001b), respectively.}
\label{circ}
\end{figure}

Those familiar with the Unification Model for Seyfert Galaxies
(Antonucci 1993)\footnote{
The Unification Model for Seyfert galaxies assumes that
Seyfert 1s (in which both Broad and Narrow lines are
visible in the optical spectrum) and Seyfert 2s 
(where only Narrow lines are observed) are intrinsically the same. The 
nucleus (i.e. the Black Hole with the accretion disc and the Broad
Line regions) is surrounded by optically and geometrically thick matter,
probably axisymmetric (usually called the `torus'). If the line--of--sight
does not intercept the torus, the nucleus is visible and the source is
classified as type 1. Otherwise, only the Narrow Line regions are visible
and the source is classified as type 2. In this scenario, a 
one--to--one relation between optical type 1 and X--ray unobscured
sources, and between type 2 and X--ray obscured sources, is expected.}
 have probably noted that I have used
the term `obscured' instead of the most common term `type 2'. 
The reason is that there is increasing evidence that optical classifications
(from which the terms type 1 and 2 derives) and X--ray classifications 
sometimes disagree with what expected from the Unification Model; 
in fact a number of sources 
clearly obscured in X--rays are either of type 1, on one extreme, or dull,
on the other extreme, when observed in the optical. I
will discuss this point in greater detailed in Sec.3. To
avoid confusion, I will confine here the terms type 1 and type 2 to their
original meaning (based on the presence or not of broad permitted lines
in the optical spectrum). In any case, I will give for granted, in agreement
with the Unification Model, that the basic X--ray properties of unobscured and
obscured AGN are the same, the latter being simply seen through a screen
of cold matter.

\section{The properties of circumnuclear matter}

\subsection{Physical properties of reflecting matter}

The advent of high spectral and spatial resolution X--ray instruments on board
$Chandra$ and XMM--$Newton$ is now allowing the study of
the circumnuclear matter
in obscured sources in much greater detail than before (and
it is now becoming feasible to study this matter also in unobscured sources,
despite the heavy dilution by the nuclear radiation). The results obtained
by these instruments are improving and refining, but not
revolutionizing, the previous scenario (for the relief and pleasure of 
those people who in the past struggled to find a coherent picture out 
of much poorer data). In Fig.~\ref{circ} the differences in the line
spectrum when observed with a CCD and with a grating instrument are
shown in the case of the Circinus Galaxy.

The spectrum of the reflected component depends on the ionization state
of the matter. If the matter is highly ionized, Compton scattering
is the most important process and the spectrum (at least up to 
a few tens of keV, where Compton recoil becomes important)
is very similar to the primary one. If instead
the matter is neutral (and optically
thick), the resulting spectrum is the so called `Compton reflection' 
with a broad bump between 10
and 100 keV (e.g. George \& Fabian 1991; Matt et al. 1991).   
In both cases strong emission lines are also expected: iron lines
from He-- and H--like ions may be present
in the former case; a EW$\sim$1--2 keV 6.4 keV iron line
is present in the latter case (e.g. Matt et al. 1996b). 
More complex continuum and line spectra are expected for mildly
ionized material.

The study of the circumnuclear matter is easier in Compton--thick sources,
just because the nucleus is completely obscured up to at least 10 keV, i.e. in
the band where imaging and high resolution spectroscopic instruments
work.  Emission from off--nuclear regions is the only visible in these
sources, and may be studied in great detail. The number and complexity
of these regions are much different from source to source. In the
Circinus Galaxy, the results obtained by Bianchi et al. (2001) from $ASCA$
and $BeppoSAX$ data have been basically confirmed by $Chandra$ (Sambruna et al.
2001a,b; see also Fig~\ref{circ}): the spectrum down to about 2 keV is dominated by one, optically
thick and with low ionization reflecting region (unresolved even
with $Chandra$), most likely the inner
surface of the $N_H\sim4\times10^{24}$ cm$^{-2}$
absorbing matter (Matt et al. 1999b) assuming, as customary, that the latter
is (in the first approximation at least) axysimmetric. A second, ionized
component extended over about 50 pc becomes important below 2 keV, where
it provides about half of the flux (Sambruna et al. 2001a,b).

The continuum and line spectra are instead much more complex in NGC~1068.
Even if significant extended emission has been observed by $Chandra$, most
of the emission is unresolved (Young et at. 2001). 
$ASCA$ and $BeppoSAX$ data already indicated the presence of more than 
one reflecting regions. Bianchi et al. (2001; see also references
therein) have indeed shown that at
least three different regions are needed to explain the line spectrum: 
the first is optically thin and moderately ionized, and it is responsible for
the  K$\alpha$ lines of elements like Mg, Si and S;
the second one is optically thin but highly ionized, and responsible
for the He-- and H--like Fe K$\alpha$ lines; 
the third one is optically thick and of low ionization, and it is responsible
for the 6.4 keV Fe and the O {\sc vii} K$\alpha$ lines. The last region
is again likely to be the inner surface of the absorber, which in this source
has a column density $\geq10^{25}$ cm$^{-2}$ (Matt et al. 1997).
XMM--$Newton$ grating spectra
(e.g. Behar et al. 2002; Kinkhabwala et al. 2002) 
have definitely proved that the emitting gas
responsible for the line spectrum 
is in photoionization equilibrium, the emission lines being due to both
recombination and resonant scattering (as predicted by e.g. Band et al. 1990;
Krolik \& Kriss 1995; Matt et al. 1996b), and are suggesting that the 
scenario is even more complex than deduced by low resolution spectra.
Photoionization equilibrium seems also able to explain the 
$Chandra$--HETG line spectrum of Mrk~3
(Sako et al. 2000). In the latter source the softest part (i.e.
below about 3 keV) of the spectrum is spatially extended along the [O III]
ionization cone, while the high--energy spectrum is unresolved and 
consistent with reflection by cold and optically thick material, once
more to be associated with the $N_H\sim10^{24}$ cm$^{-2}$ absorber
(Cappi et al. 1999). 

Combining these results with those discussed by Matt et al. (2000), who
studied a sample of bright Compton--thick sources observed by $BeppoSAX$,
it is possible to conclude 
that reprocessing from optically thick, almost neutral
matter is quite common. As said above, 
it seems natural to identify this matter with
the inner surface of the absorber, as all these sources are Compton--thick.
(We will see later that this matter is likely to be present also in most
unobscured sources, according to the Unification Model, 
but even in Compton--thin obscured sources, which is less obvious). 
In order not to exceed the dynamical mass, at least in 
Circinus and NGC~1068 the matter must be fairly close to the 
black hole, within a few tens of pc at most (Risaliti et al. 1999). 
(A completely independent estimate of the inner radius 
has been obtained by
Bianchi et al. 2001 for these two sources, by modeling the X--ray line spectra.
They found a minimum distance of the torus of about 0.2 and 4 pc,
respectively). Indeed, in the Circinus Galaxy the nuclear cold reflector
is unresolved, which implies un upper limit to its size of about 15 pc
(Sambruna et al. 2001a). 

\subsection{Compton--thin or compton--thick?}

\begin{figure}
\epsfig{file=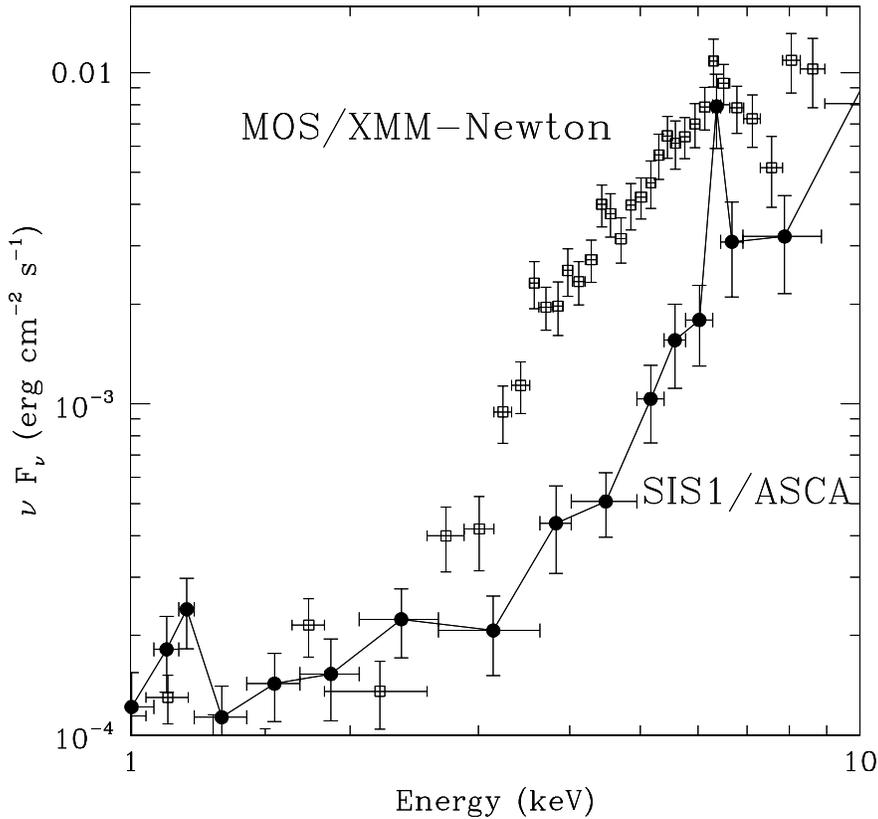, height=12.0cm, width=12.0cm}
\caption{The ASCA and XMM--$Newton$ spectra of UGC4203. From Guainazzi 
et al. (2002).}
\label{phoenix}
\end{figure}

Unless the nucleus is directly observed above $\sim$10 keV (a task
possible in recent years with $BeppoSAX$ and $RXTE$, 
and only for column densities
not exceeding $\sim$10$^{25}$ cm$^{-2}$, see Matt et al. 2000), 
the signature of Compton--thick absorption 
is a reflection dominated X--ray spectrum. Sometimes, however, a pure
reflection spectrum may lead to a wrong classification. The 
classical case is NGC~4051, which was observed by $BeppoSAX$
in a low flux state, in which the nucleus was switched off and 
only the reflection component was visible (Guainazzi et al. 1998). 
The source was a well known Seyfert 1, otherwise it would have
been classified as a Compton--thick absorbed AGN. 
A similar, even if somewhat less dramatic, change occurred in the
Compton--thin ($N_H\sim10^{22}$ cm$^{-2}$) 
source NGC~2992 (Gilli et al. 2000) which, during a 
$BeppoSAX$ observation, was almost switched off, with the reflection 
component thus becoming very prominent. Subsequent $BeppoSAX$ observations
of NGC~4051 and  NGC~2992
found that both sources had recovered their normal, bright state.
In both cases it seems pretty obvious that what changed was the nuclear
flux, rather than the properties of the absorption. It is also clear that
the reflecting matter must be located at large distances from the black
hole to echo the already switched--off primary emission; in fact, reflection
from the accretion disc would disappear almost immediately. 
It is worth noting
that, in the case of NGC~2992, the absorbing matter (which is Compton--thin)
must be different from the reflecting matter (which is Compton--thick). 
Because the galaxy is edge--on, and given the rather small
column density, it is possible that the thin absorber
is the disc of the  host galaxy. 

Recently, two more sources underwent a similar transition.
NGC~6300 and UGC~4203 were both Compton--thick when observed by
RXTE (Leighly et al. 1999) and $ASCA$ (Awaki et al. 2000), respectively,
but become Compton--thin when observed later on by $BeppoSAX$ (Guainazzi 2002)
and XMM--$Newton$ (Guainazzi et al. 2002). 
The $ASCA$ and XMM--$Newton$ spectra of UGC~4203
are presented in Fig.~\ref{phoenix}. While a change in the properties
of the absorber cannot be completely ruled out, 
the explanation in terms of a `switching--off' of the sources during their
first observation seems the most natural. The `thin' absorbers in these
two sources have column densities of 
a few$\times$10$^{23}$ cm$^{-2}$, and the host galaxies are
seen at low inclination angles (UGC~4203 is basically face--on), so
the galactic disc cannot be the cause for the Compton--thin absorption. 
This therefore strongly suggests that in these two sources both 
Compton--thin (the absorber) and Compton--thick (the reflecting) 
cold, circumnuclear regions are present. 

\subsection{Two different regions?} 

Because X--ray absorption is very common in AGN, the covering factor
of the absorbing matter must be large. Heavy absorption is also
very common: about half of the optically selected Seyfert 2s in the local
Universe are Compton--thick (e.g. Maiolino et al. 1998). Indeed, the very
first object observed by XMM--$Newton$ in the framework of a program devoted
to study the absorption properties of optically selected Seyfert 2s, NGC~4968,
is clearly a Compton--thick source (Guainazzi et al., in preparation). 
Even allowing 
for a number of misclassifications due to the switching--off of the nucleus
rather than a true Compton--thick absorption, as discussed in the previous
paragraph, it is clear that optically thick circumnuclear matter is common. 

Risaliti et al. (1999) have shown that there is a relation between optical
classification and column density, at least for optically selected
AGN: Intermediate (1.8-1.9) Seyferts are usually Compton--thin,
while classical Seyfert 2s are Compton--thick. 
From the results discussed in the previous paragraph, it seems that
the two components may be present in the same source (another
case of Compton--thin absorber with Compton--thick reflection is NGC~5506,
Matt et al. 2001). 

All these pieces of evidence suggests that the Compton--thin and
Compton--thick materials have different origins and locations, as originally
proposed by Maiolino \& Rieke (1995). Matt (2000) has suggested
that the Compton--thin matter should be associated with the dust lanes
at distances of hundred of parsecs which Malkan et al. (1998) found to be
common in Seyfert galaxies, while the Compton--thick matter should 
be much closer to the nucleus, and associated with the `torus' 
envisaged in the Unification Model (Antonucci 1993). A similar scenario
has been proposed by Weaver (2002), who identifies the Compton--thin
absorbers with starburst region clouds. In any case it is clear
that the Unification Model should be somewhat revised to accomodate this
further component. Other problems with the 
Unification Model have emerged from recent X--ray surveys and will be
discussed in the next section.

\section{Obscured AGN and the Unification Model}

\begin{figure}
\epsfig{file=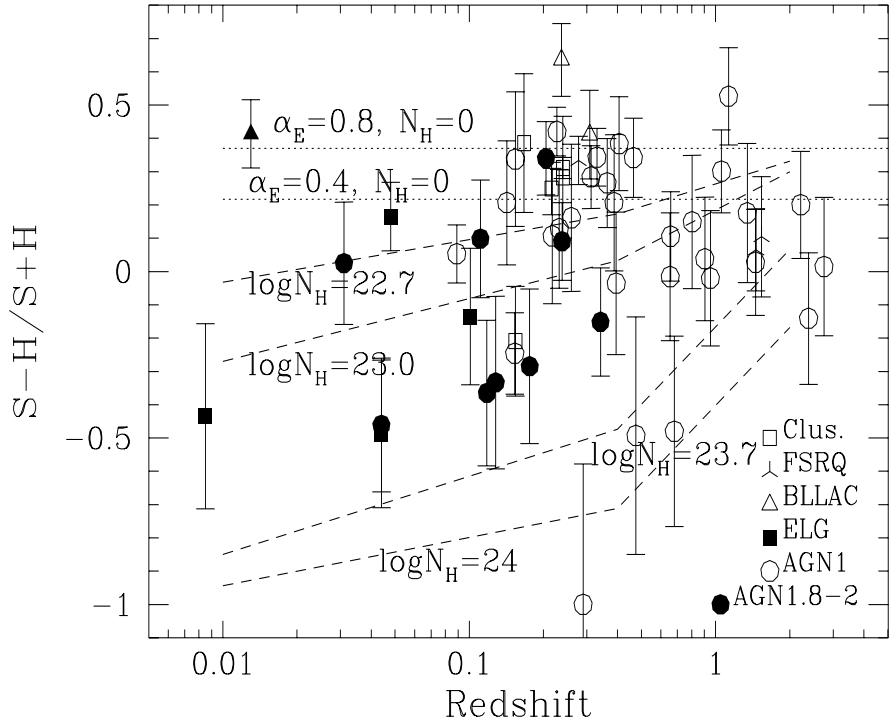, height=14.0cm, width=12.0cm, angle=-90}
\caption{The softness ratio (S-H)/(S+H) as a function of redshift for
the identified BeppoSAX {\sc hellas} sources. S and H are the counts in
the 1.3--4.5 and 4.5--10 keV bands, respectively. Note that there are
a number of absorbed type 1 AGN. From Comastri et al. (2001).}
\label{hellas}
\end{figure}

A first (actually not very serious) problem for the Unification Model
 has been already
mentioned: the co--existence of two kinds of cold, circumnuclear matter,
which suggests that the torus (i.e. the Compton--thick matter) is not the
only possible absorbing/reflecting region. The inclusion of dust lanes 
and/or starburst regions to the basic ingredients should be all what is needed
to explain the observations. 

There are, however, more serious problems. First of all,
type 1 sources may be obscured in X--rays. Maiolino et al. (2001a)
found a number of broad lines QSOs with significant X--ray absorption. Several
sources in e.g. the BeppoSAX {\sc hellas} (Fiore et al. 2001), the
ASCA HSS (Della Ceca et al. 2001) and the XMM--$Newton$ 
{\sc hellas2xmm} (Fiore et al. 2002) surveys appear to be 
absorbed in X--rays, but optically identified with type 1 AGN
(see Fig.~\ref{hellas}). Recently, it has been shown
(e.g. Gallagher et al. 2002; Mathur et al. 2001) that the X--ray weakness
of BAL QSOs discovered by ROSAT is due to excess absorption (but it is
still not clear whether the X--ray absorption is associated with the UV
one, and therefore significantly ionized). 
For these absorbed blue QSOs, and in general for
all the AGN for which optical extinction is lower than expected from
X--ray absorption (there are several
cases among Seyfert 2s), a possible explanation is that the dust/gas
ratio of the absorber is much lower than in the Galactic interstellar
medium, probably due to dust sublimation. Another possibility is that
the optical and X--ray absorbers  cover different regions: the
very nucleus for the X--ray absorber, a larger region for the optical one.
Recently, Maiolino et al (2001b) suggested an explanation in terms of different
dust grain size with respect to the ISM. Whatever the solution, 
it is clear that optical and X--ray observations may sometimes lead 
to different classifications. 

There are also X--ray loud sources (and sometimes, but not
always, with hard spectra, thus suggesting obscuration) that
do not appear as AGN in the optical (or even do not have an optical
counterpart, implying a very large X--to--optical ratio). Such a population 
is starting to emerge at fluxes of about 10$^{-14}$ erg cm$^{-2}$ s$^{-1}$
(e.g. Mushotzky et al. 2000;
Fiore et al. 2001; Barger et al. 2001; Alexander et al. 2001),
now easily probed by  $Chandra$ and XMM--$Newton$.
The nature of these X--ray Bright, Optically Normal Galaxies (XBONG,
as they were christened by Comastri et al. 2002a, where a
detailed discussion of the properties of 10 {\it bona fide} sources of
this class can be found) is still matter of debate. A large fraction
of these sources have X--ray luminosities exceedings 10$^{42}$ erg s$^{-1}$,
well in the AGN regime. My personal feeling is that
most, if not all, of them
will turn out to be obscured AGN, in which the obscuration is such to
prevent even the formation (or the visibility)  of the Narrow lines.
Compton--thick obscuration may explain the not particularly hard spectrum
of some of them, if reflection from ionized matter is not negligible with
respect to reflection from cold matter (with an appropriate choice of the
bands, even the archetypal Compton--thick source, 
NGC~1068, would appear as a very soft source indeed!)
It is worth recalling that 
cases of obscured AGN whose optical spectrum, even if not really `normal', is 
nevertheless different from that of an AGN are already known: NGC~6240 and
NGC~4945 are the best examples, see Matt (2002) and references therein.

The most extensively studied among XBONG is probably
`P3', one of the sources in the sample of Fiore et al. (2000). 
This object was observed in radio (ATCA), near--IR (ISAAC/VLT), 
optical (ESO 3.6m) and X--rays (both $Chandra$ and XMM--$Newton$),
and discussed in Comastri et al. (2002b). The
X-ray luminosity (a few times 10$^{42}$ erg s$^{-1}$ if the source
is Compton--thin; $\simgt$10$^{44}$ erg s$^{-1}$ if Compton--thick) 
and hard X--ray spectrum clearly indicate the presence of an AGN
for which, however, there are no signatures at longer wavelengths.
The analysis of the SED makes the ADAF solution rather unlikely. An heavy
obscured AGN appears to be the most likely explanation, even if a rather
extreme BL Lac cannot be ruled out.

Of course, what just said does not mean that 
there are no relations whatsoever between optical and X--ray
appearances. More often than not, the optical (X--ray) appearance is just
what one would predict from the Unification Model  after observing
the X--ray (optical) emission. Moreover, we are
not aware of any certainly unobscured AGN which are not
type 1,  and of any type 2 which are not obscured (even if Pappa
et al. 2001 presented a couple of possible cases). 
So, at present the `strict' (in the sense of no unambiguous 
exceptions found yet) relations between optical and X--ray 
classifications may be reduced to:

\begin{center}
{\bf type 1 $\leftarrow$ unobscured}
\end{center}

\begin{center}
{\bf type 2 $\rightarrow$ obscured}
\end{center}

It is worth noting that support for one of the predictions of 
the Unification Model, i.e. that Seyfert 1 galaxies have the `torus' (and therefore
would become Seyfert 2 if observed at different angles) is coming
from recent $Chandra$ and XMM--$Newton$ observations, which
are indicating that the presence of narrow iron K$\alpha$ lines
(therefore produced in distant matter; iron lines from the
accretion disc being expected to be broad due to kinematic and
relativistic effects, see Fabian, this volume) are rather common
(e.g.: Yaqoob et al. 2002, Weaver 2002  and references therein for
$Chandra$ results; Matt et al. 2001, Pounds \& Reeves 2002 and
 references therein for XMM--$Newton$ results).

\section{Obscured AGN and the X-ray and IR Cosmic Backgrounds}

Obscured AGN are a basic ingredient in synthesis models for
the CXRB (e.g. Setti \& Woltjer 1989; Comastri et al. 1995). The mixture
of unobscured and obscured (with a spread of column densities) AGN is
able to reproduce well the spectral shape of the CXRB, so solving the
long standing problem known as the 
`spectral paradox'. A large fraction of the CXRB below 10 keV
has been now resolved in discrete sources, many of them optically identified
as AGN (and part of them, as said above, {\it assumed} to be AGN due to their
X--ray luminosity). Some problems are still to be
solved (see talks by Hasinger and Brandt in this volume) but, after
40 years from its discovery, the main issue, i.e. what class
of sources make the CXRB (or at least most of it), can be now considered
settled (and this despite that the
bulk of the CXRB, which peaks around 30 keV, is still far to be resolved
for lack of sensitive imaging  instruments in that band). 

Cosmic backgrounds have been recently measured also in other
bands, notably the sub-mm and IR.
When looking at the Spectral Energy Distribution of the extragalactic
Backgrounds, the CXRB may appear almost negligible if compared with
the cosmic IR background (CIRB). However,
the luminosity we observe in the CXRB is only a fraction,
probably of the order of 10-20\%, of the energy actually emitted, the
remaining flux having been absorbed by the circumnuclear matter, 
and re--emitted at longer wavelengths, mostly in the mid--IR. 
Fabian \& Iwasawa (1999) estimated that a by no means
negligible fraction (several tens per cent) of the CIRB is
actually due to X--ray photons absorbed and reprocessed from the obscuring
medium (see also  Risaliti et al. 2002). Indeed, by
cross--correlating IR and X--ray deep observations 
Fadda et al. (2002) found an AGN contribution of around 15\% of the
CIRB at 15$\mu$.
From optical spectroscopic identifications of ISO/ELAIS sources, Matute et al.
(2002) estimate a contribution at 15$\mu$ of 10-15\%. These two estimates
are both likely to be lower limits, as in the former case some Compton--thick
sources may have been missed, and in the latter case AGN which 
are bright in IR but do not show AGN--like optical lines,
like NGC~6240 and NGC~4945 (Matt 2002) may have
not been recognized as AGN. 

Whatever the real number is, it is clear that accretion and star
formation are processes  of comparable importance
in the history of the Universe.

\medskip
\noindent
{\it Acknowledgements} It is a pleasure to thank all my collaborators
in these researches: S. Bianchi, A.C. Fabian, F. Fiore, M. Guainazzi,
K. Iwasawa, G.C. Perola and all members of the {\sc hellas} team.
I acknowledge financial
support from ASI and MURST (under grant {\sc cofin-00-02-36}).

\end{document}